\documentclass[mathleft,fleqn]{an}
\usepackage{amsmath}
\usepackage{graphicx}
\usepackage{ulem}
\usepackage{natbib}
\bibpunct{(}{)}{;}{a}{}{,}
\newcommand{\dnu}{\left\langle \Delta \nu \right\rangle}
\newcommand{\dnusol}{\left\langle \Delta \nu  \right\rangle _{\odot}}
\newcommand{\numax}{\nu _{\rm max}}
\newcommand{\numaxsol}{\nu _{\rm max, \odot}}
\newcommand{\edit}[1]{{#1}}

\usepackage{placeins}

\begin{document}


\Pagespan{1}{}
\Yearpublication{2015}%
\Yearsubmission{2015}%
\Month{12}%
\Volume{999}%
\Issue{0}%
\DOI{asna.201400000}%

\title{Estimating the ages of red giants with Asteroseismology}
\title{Asteroseismology of red giants: from analysing light curves to estimating ages}
\title{A how-to compendium on  asteroseismology of red giants}
\title{From analysing light curves to estimating stellar ages: a compendium of asteroseismic methods}
\title{From analysing light curves to estimating stellar ages: a compendium on asteroseismology of red giants}
\title{A compendium on asteroseismology of red giants: from analysing light curves to estimating ages}
\title{Asteroseismology of red giants: from analysing light curves to estimating ages}


\author{G.\,R.\,Davies\inst{1}\fnmsep\thanks{Corresponding author:
        {grd349@gmail.com}} \and A.\,Miglio\inst{1}
}
\titlerunning{Estimating the ages of red giants with Asteroseismology}
\authorrunning{G.\,R.\,Davies and A.\,Miglio}
\institute{
School of Physics and Astronomy, The University of Birmingham, Edgbaston, B15 2TT, UK.
}

\received{XXXX}
\accepted{XXXX}
\publonline{XXXX}


\abstract{
Asteroseismology has started to provide constraints on stellar properties that will be essential to accurately  reconstruct the history of the Milky Way.  Here we look at the information content in data sets representing current and future space missions (CoRoT, {\it Kepler}, K2, TESS, and PLATO) for red giant stars.  We describe techniques for extracting the information in the frequency power spectrum and apply these techniques to {\it Kepler} data sets of different observing length to represent the different space missions.  We demonstrate that for KIC 12008916, a low-luminosity red giant branch star, we can extract useful information from all data sets, and for all but the shortest data set we obtain good constraint on the g-mode period spacing and core rotation rates.  We discuss how the high precision in these parameters will constrain the stellar properties of stellar radius, distance, mass and age.  We show that high precision can be achieved in mass and hence age when values of the g-mode period spacing are available. We caution that tests to establish the accuracy of asteroseismic masses and ages are still ``work in progress''.}

\maketitle

\section{Introduction}
\begin{figure}
\includegraphics[width=0.98\linewidth]{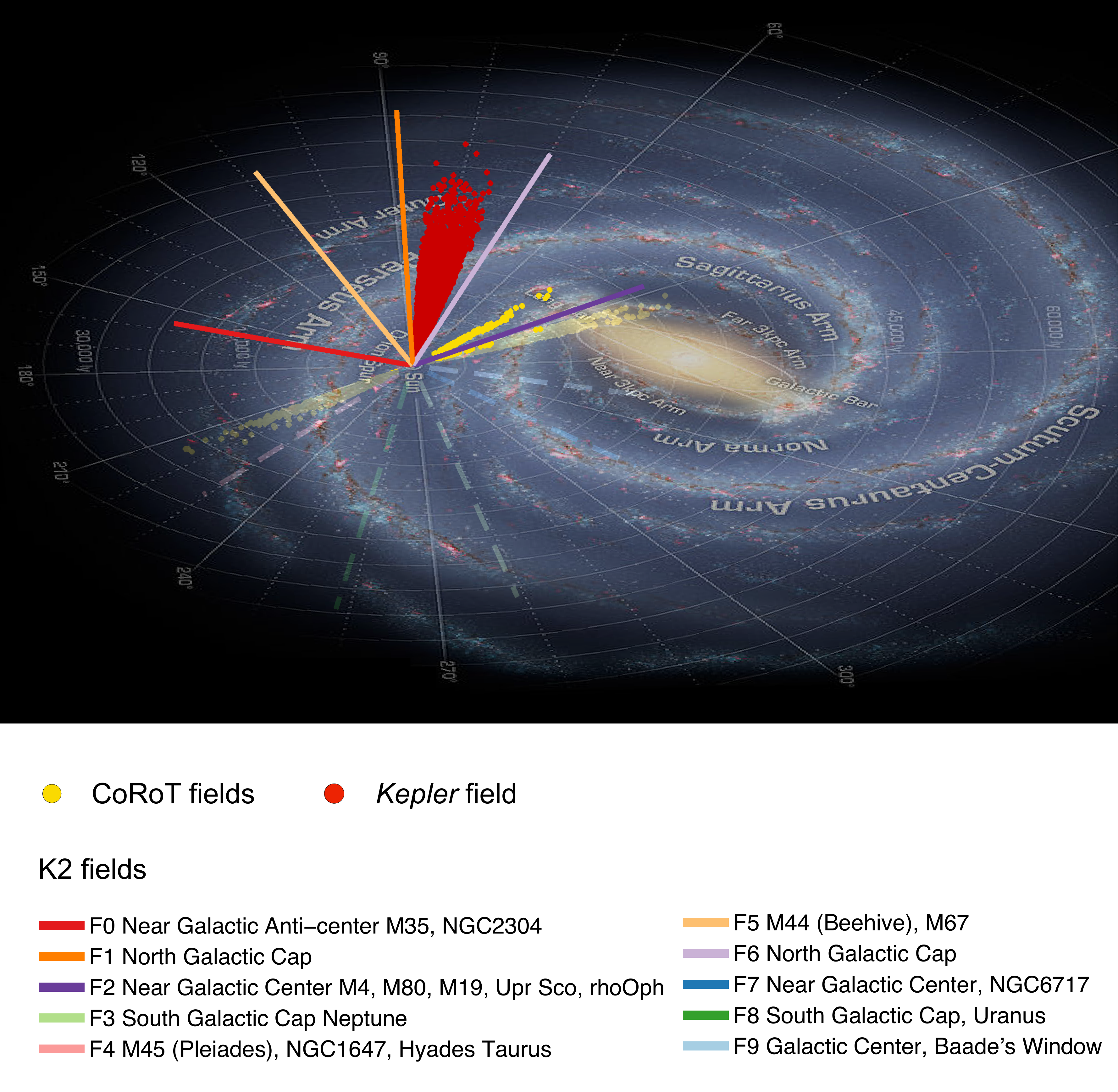} 
\caption{An artist's conception of our Milky Way galaxy (CREDIT: NASA/JPL-Caltech  http://www.jpl.nasa.gov/news/news.php?release=2010-179) with fields of view for {\it Kepler}, CoRoT, and K2.}
\label{fig::dir}
\end{figure}

In recent years asteroseismology has entered its golden age.  With the advent of the {\it Kepler} \citep{2010Sci...327..977B} and CoRoT \citep{2006ESASP.624E..34B} space missions, time series measuring stellar variability of very-high quality have become widely available.  Analysis of these time series can deliver precise estimates of stellar ages \citep{Lebreton2014,2015MNRAS.452.2127S, 2015ApJ...811L..37M, 2015MNRAS.446.2959D, 2015Natur.517..589M, Miglio2013a} - a quantity critical for reconstructing the history of the Milky Way.  With the re-purposed {\it Kepler} mission K2 \citep{2014PASP..126..398H} currently capturing solar-like oscillations \citep{2015arXiv150701827C,2015ApJ...809L...3S} in a number of different galactic directions, and the future missions of TESS \citep{2014SPIE.9143E..20R} and PLATO \citep{2014ExA....38..249R} adding to this, ages for many thousands of stars in many different galactic distances and directions present an exciting possibility.  Figure \ref{fig::dir} shows fields of view for the {\it Kepler}, CoRoT, and K2 space missions.
\par  
One method of using variability to estimate stellar age is asteroseismology.  In this work we restrict our discussions to the asteroseismology of solar-like pulsators. Solar-like pulsators are a class defined by the presence of surface convection.  This turbulent convection excites (and intrinsically damps) modes of acoustic oscillation.  The modes of oscillation can propagate throughout the star.  Because mode frequencies are sensitive to the size of the cavity and the sound speed in the cavity, oscillation modes are probes of the structure of the stellar interior.  For more details see \cite{2013ARA&A..51..353C}.
\par  
In this proceedings we discuss the measurement of seismic parameters from frequency power spectra and how these observables can provide constraint on inferred stellar properties.  We start by considering the power spectrum for an example star and discuss how to reduce the data to observables that can be compared with stellar models.  We finish by showing examples of the constraint on stellar mass and age that these observables provide.  
\section{The power spectrum}
\label{sec::defs}
Here we present results of a power spectrum modelling procedure applied to a single section of the power spectrum of KIC 12008916.  This star is a low-luminosity red giant and exhibits an exquisite pattern of mixed mode oscillations in the Kepler observations spanning Q0 through to Q17.
\par
\begin{figure*}
\includegraphics[width=0.48\linewidth]{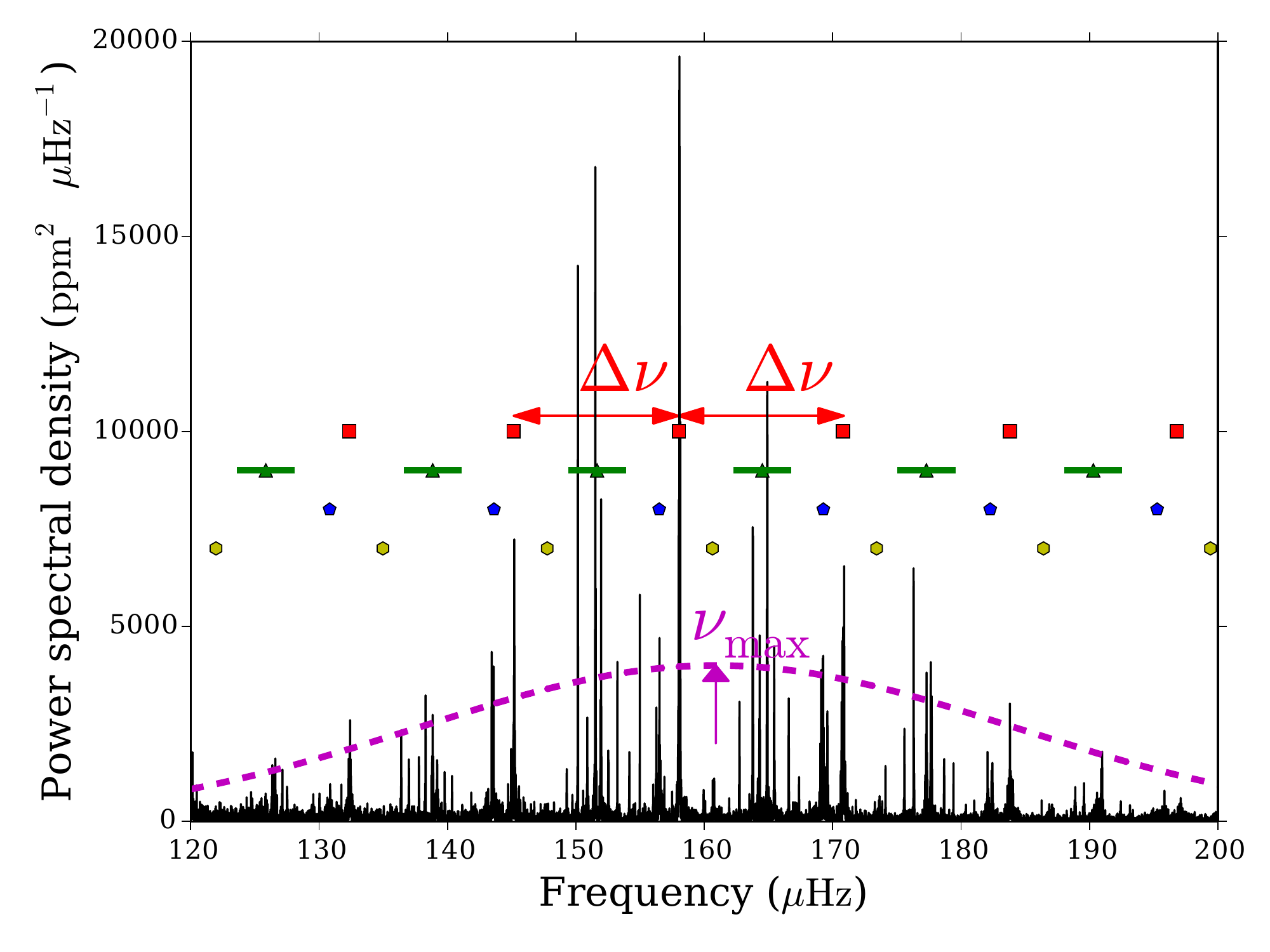}
\includegraphics[width=0.48\linewidth]{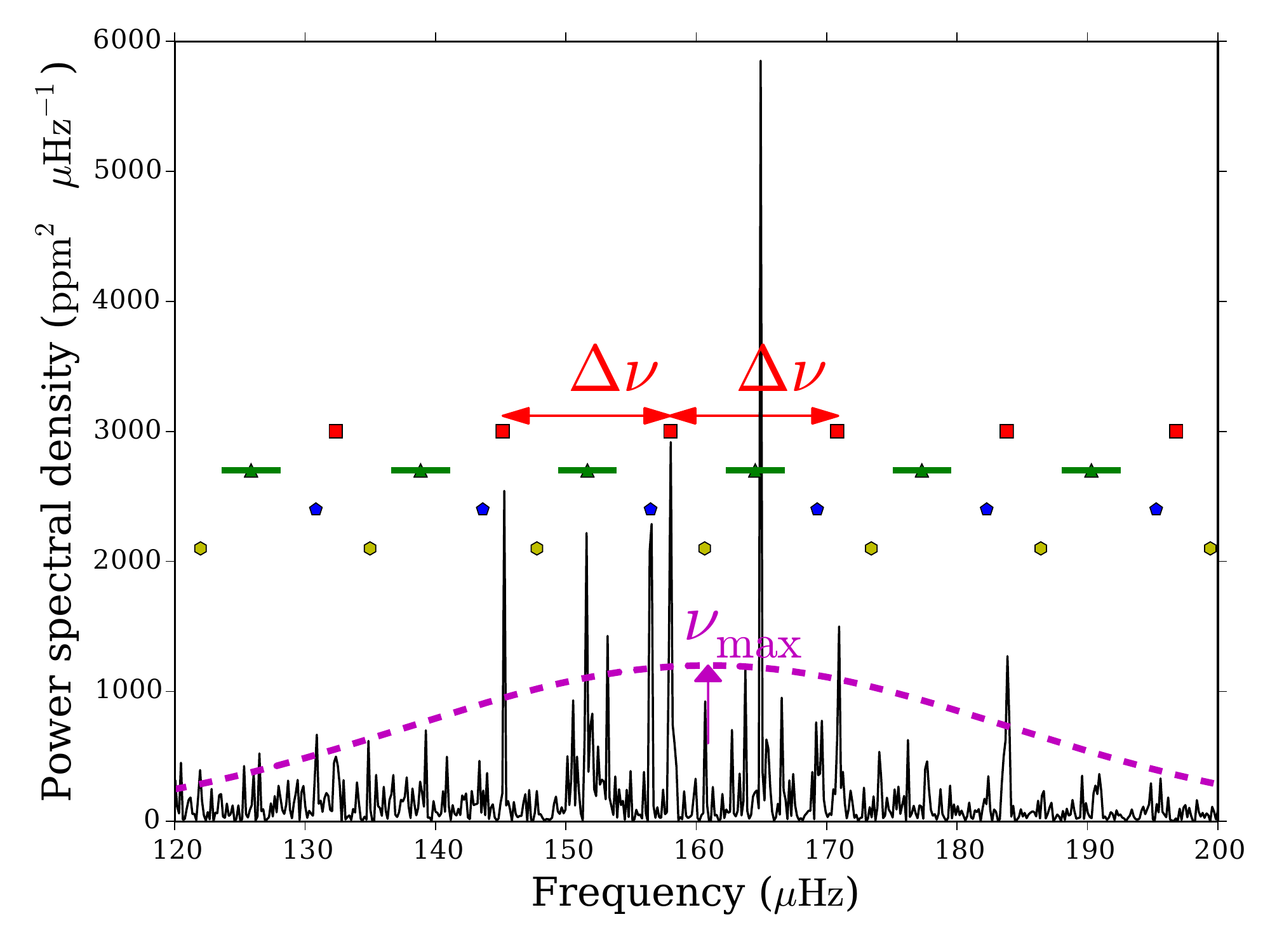}
\caption{Power spectra for KIC 12008916.  Left: length of data set 1335 days.  Right: length of data set 70 days.  We have illustrated the plot with mode frequencies and identifications, together with the global properties.  Symbols represent the radial modes (red squares), the dipole modes (green triangles with extended bars), quadrupole modes (blue pentagons), and octupole modes (yellow hexagons).  We have shown examples of the large separation ($\Delta \nu$) between radial modes and have illustrated the oscillation envelope of power, the maximum of which is labelled as $\nu_{\rm max}$.}
\label{fig::psd}
\end{figure*}

\begin{figure*}
\includegraphics[width=0.48\linewidth]{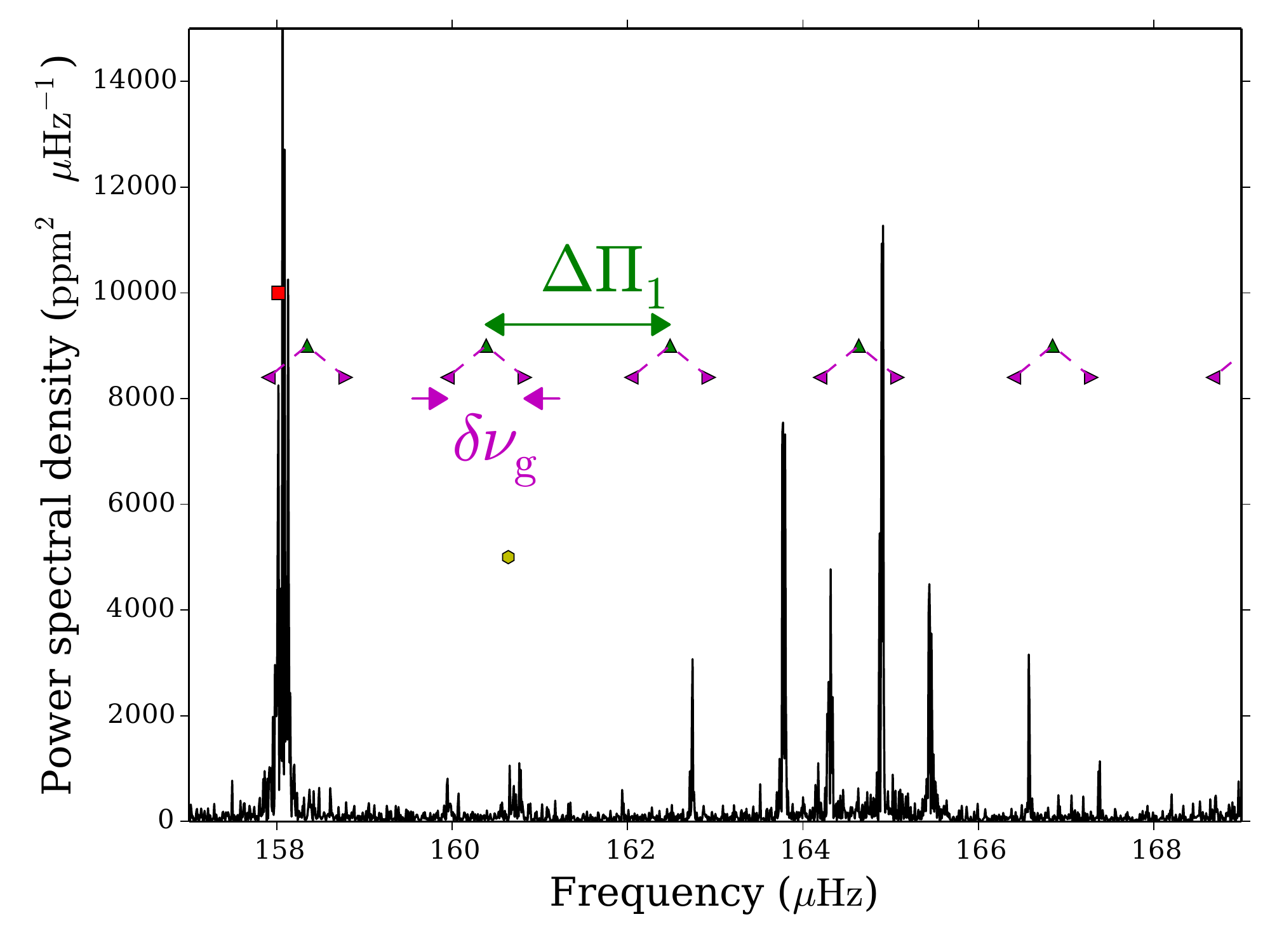} 
\includegraphics[width=0.48\linewidth]{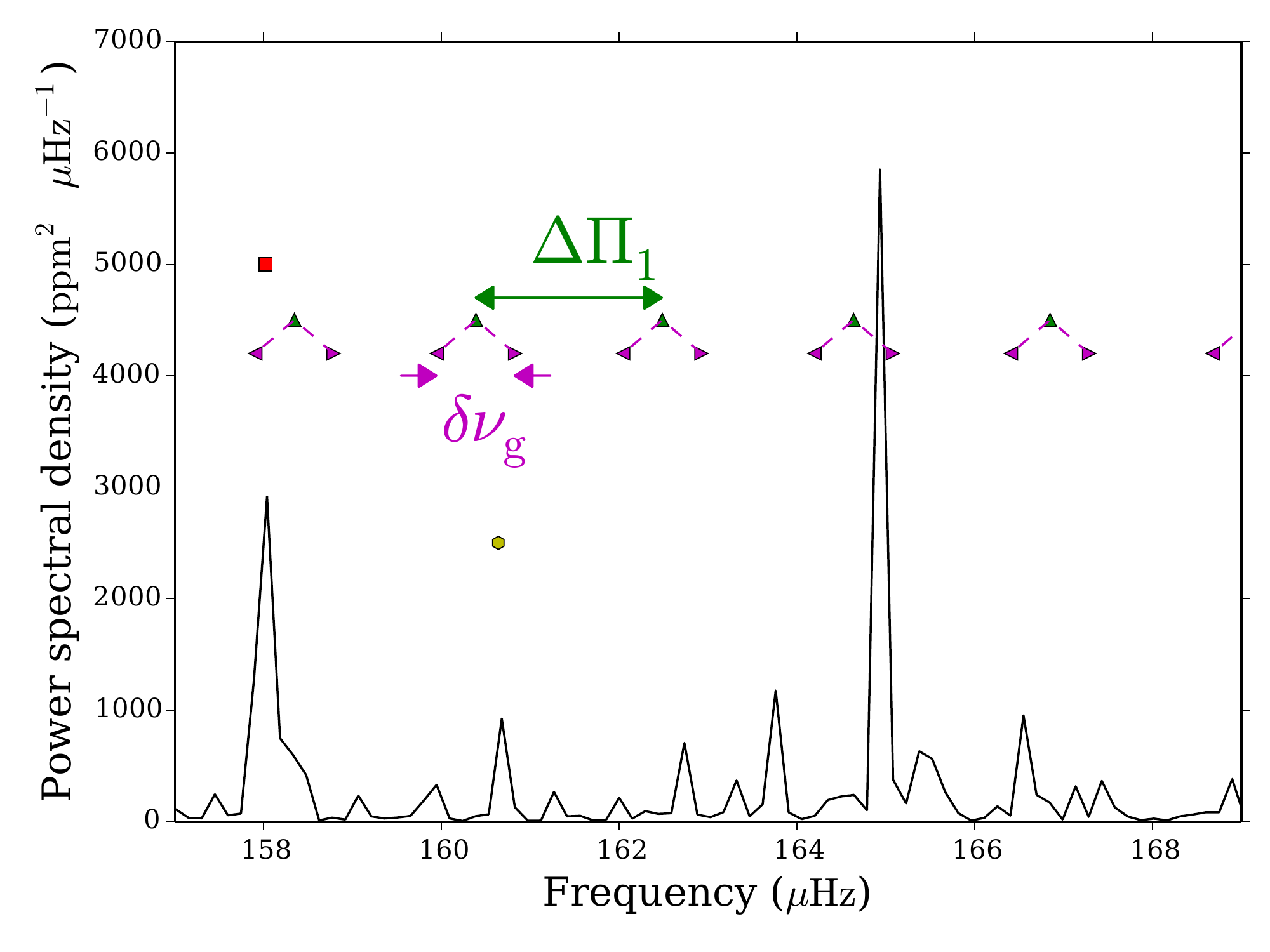}
\caption{Power spectra for KIC 12008916.  Left: length of data set 1335 days.  Right: length of data set 70 days.  We have illustrated the plot with g-mode dipole frequencies (green up triangles), rotational splitting (magenta left or right triangles), the radial p-mode frequency (red square), and the octupole p-mode frequency (yellow hexagon).  We have shown an example of the g-mode period spacing and the core rotational splitting parameter.  Note that we do not observe simple g modes but in fact observe a more complex pattern of modes with mixed p and g character.}
\label{fig::zoom}
\end{figure*}

 
The light curve was prepared from public data available through the KASOC website.  The raw light curve was detrended using a simple smoothing algorithm and then the power spectrum was computed following the procedure of \cite{2011MNRAS.414L...6G}.  The mode power spectrum, and the section in frequency we fitted our detailed model to, can be found in Figures \ref{fig::psd} and \ref{fig::zoom}.  Marked in these plots are examples of the properties we will consider here: the so called global properties, the frequency of maximum power $\nu_{\rm max}$ and the \edit{large frequency spacing} $\Delta \nu$; and properties related to the core, the gravity-mode period spacing of the dipole modes $\Delta \Pi_{1}$ and the frequency splitting of the gravity dipole modes as a result of core rotation $\delta \nu_{g}$.  We will see that the $\nu_{\rm max}$, $\Delta \nu$, and $\Delta \Pi_{1}$ measured in red giants can be combined to give excellent constraint on stellar age.

\par 
Our ability to extract properties of solar-like oscillations depends on a number of factors, but principally the signal-to-noise ratio (SNR) of the oscillations to background and the length and cadence of the observations.  The star we study here (KIC 12008916) exhibits a very high SNR \citep{2015A&A...579A..83C} for the best modes of oscillation ($> 250$), but detections of individual oscillations are possible at much lower SNR (typically $~ 8$).  In the absence of instrumental noise (including shot noise), the SNR of the pulsation spectrum is typically determined by the properties of the star (for pulsations see \citealt{2010ApJ...723.1607H} and background see \citealt{2014A&A...570A..41K}).  Instrumental noise is somewhat specific to each space mission but is in general a function of stellar magnitude, so that brighter stars have lower instrumental noise and hence higher SNR.
\par 
\section{Obtaining global properties of the oscillations}
The global properties of oscillation modes are commonly summarised by two parameters: the average large separation $\dnu$ and $\numax$.  With the addition of the stellar effective temperature ($T_{\rm eff}$) the mass ($M$) and radius ($R$) can be estimated from the common scaling relations (scaled to solar values):
\begin{equation}
\left(\frac{R}{{\rm R_{\odot}}} \right) \simeq \left( \frac{\numax}{\numaxsol}\right) \left(\frac{\dnu}{\dnusol} \right)^{-2} \left(\frac{T_{\rm eff}}{T_{\rm eff, \odot}} \right)^{0.5},
\label{eq::radius}
\end{equation} 
and
\begin{equation}
\left(\frac{M}{{\rm M_{\odot}}} \right) \simeq \left( \frac{\numax}{\numaxsol}\right)^{3} \left(\frac{\dnu}{\dnusol} \right)^{-4} \left(\frac{T_{\rm eff}}{T_{\rm eff, \odot}} \right)^{1.5}.
\label{eq::mass}
\end{equation} 
\subsection{Estimating $\numax$}
\begin{figure}
\includegraphics[width=\linewidth]{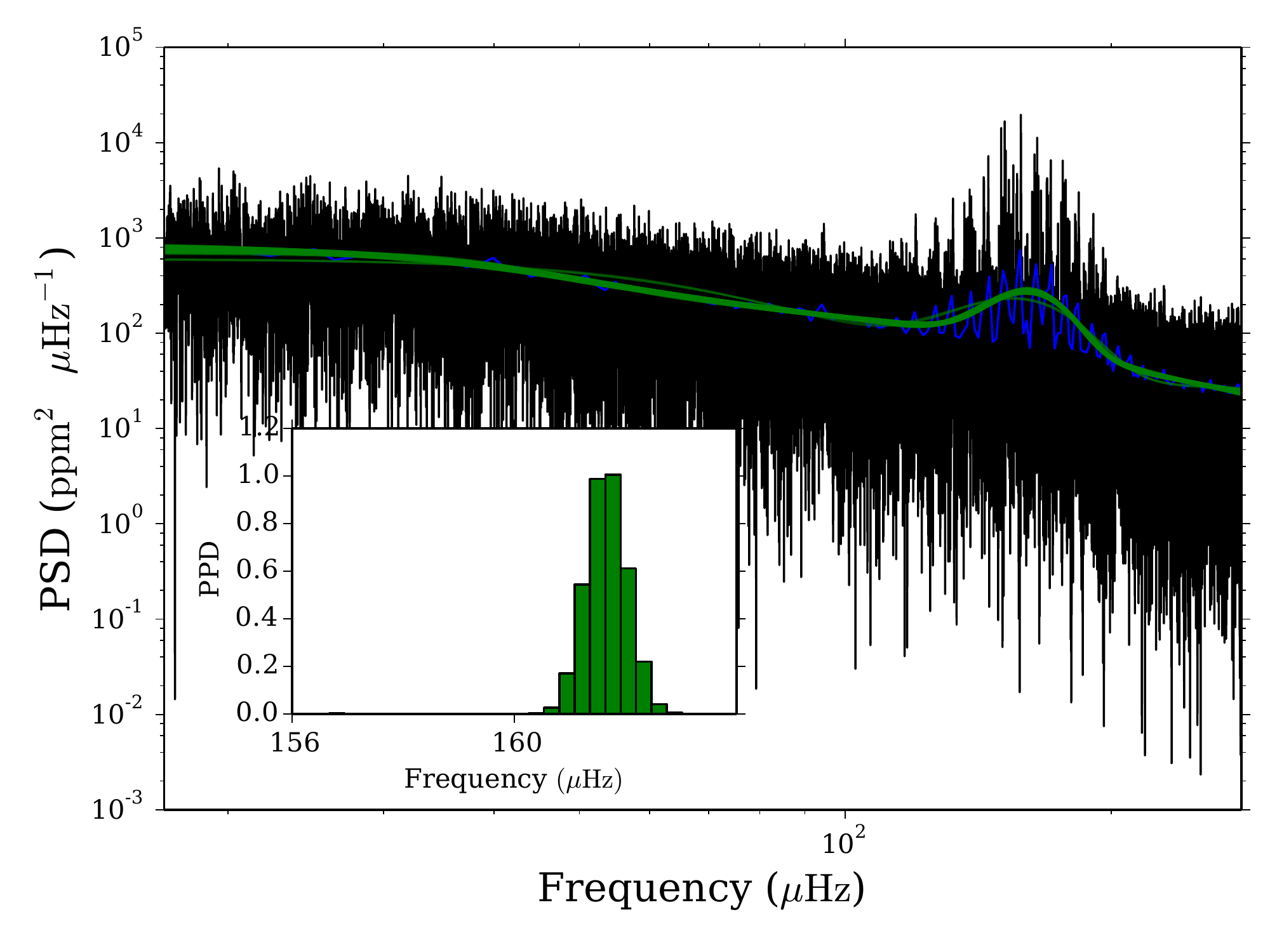} 
\caption{Background fit to the power spectrum of KIC 12008916.  Main plot: black, unsmoothed power spectrum; blue, box-car smoothed power spectrum; green, models of draws from the markov chains representing the fit to the data.  Insert: Posterior probability distribution of $\numax$.}
\label{fig::back}
\end{figure}

To estimate the frequency of maximum oscillation power we fitted a background model to the data.  We fitted model H of \cite{2014A&A...570A..41K}, comprised of two Harvey profiles, a Gaussian oscillation envelope, and an instrumental noise background.  For the estimate of $\numax$ we took the central frequency of the Gaussian component.  Figure \ref{fig::back} shows the fit to the data and the resulting marginalised posterior probability density for $\numax$.  We summarised the normal-like posterior probability with the median and the standard deviation. In this fit to the full data set we obtained $\numax = 160.9\pm0.5 \, \rm \mu Hz$. 

\subsection{Estimating $\dnu$}
\begin{figure}
\includegraphics[width=\linewidth]{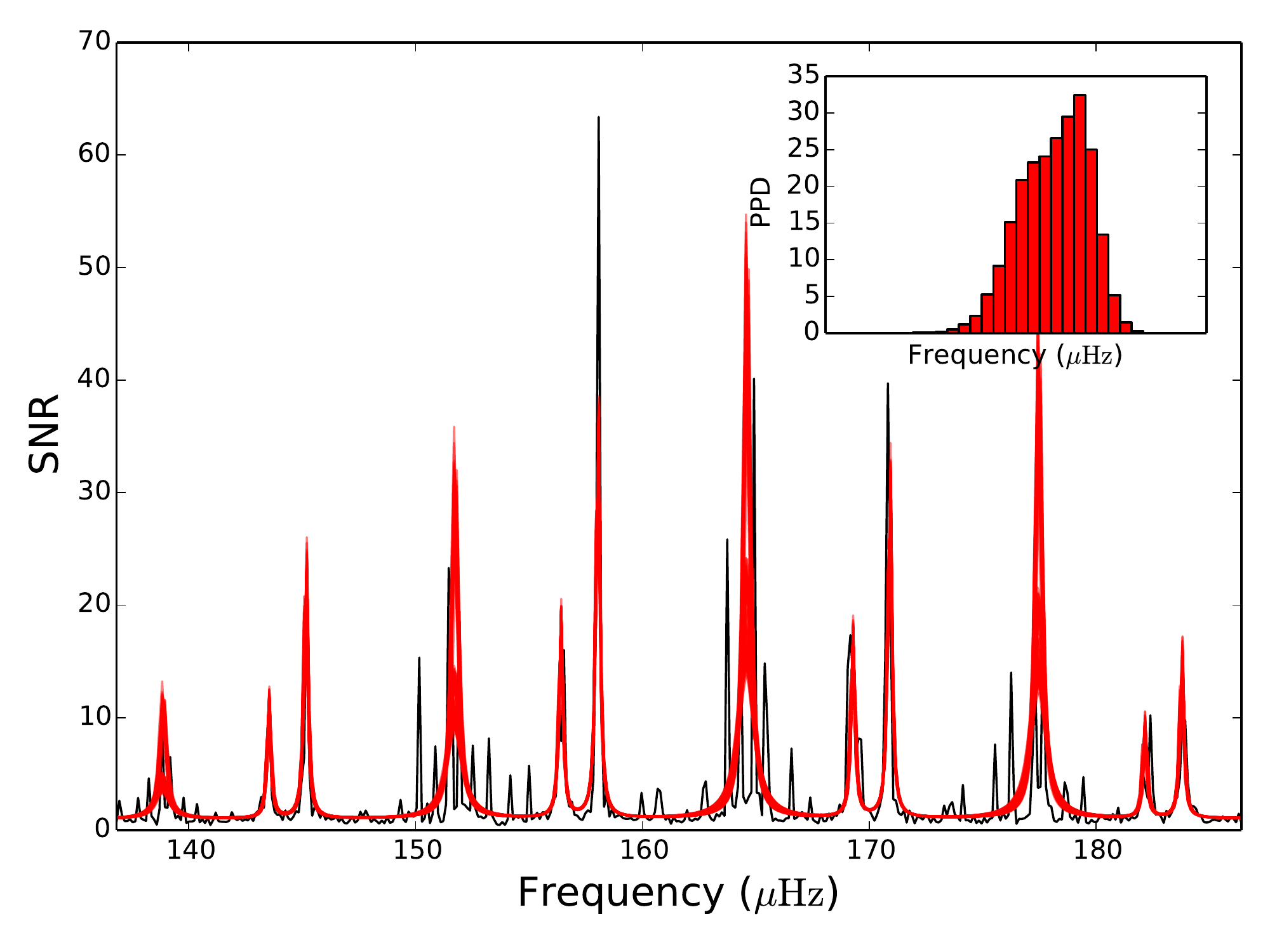} 
\caption{Radial and quadrupole mode fit to the signal-to-noise ratio spectrum of KIC 12008916.  Main plot: black, unsmoothed power spectrum; red, models of draws from the markov chains representing the fit to the data.  Insert: Posterior probability distribution of $\dnu$.}
\label{fig::dnu}
\end{figure}

To estimate the average large frequency spacing we fitted a simple model to the signal-to-noise ratio (SNR) spectrum.  We calculated the SNR spectrum by dividing the power spectrum by our background fit (with the Gaussian oscillation component suppressed) giving a noise spectrum with mean background of unity plus some modes of oscillation.  We selected a region in frequency around $\numax$ plus and minus twice an initial estimate of $\dnu$.  \edit{We then fitted a model as the sum of Lorentzian profiles that represent radial and quadrupole modes separated by some large and small frequency separations.}  The average large separation estimated was then simply the summary statistics (median and standard deviation) of the posterior probability distribution of the large spacing parameter.  Figure \ref{fig::dnu} shows the fit to the data and the posterior probability of $\dnu$.  Again, we summarise the posterior probability with the median and the standard deviation.  In this fit to the full length of the time series we obtain $\dnu = 12.89\pm0.01 \rm \, \mu Hz$.
 
\section{Detailed modelling of the power spectrum}
In order to model the detail in the power spectrum we used asymptotic expressions to determine model mode frequencies \citep{1989nos..book.....U}.  We used concepts of the mode inertia to determine the properties of each mode of oscillation.  We started by describing a model of the power spectrum from a range in frequency that incorporates the radial mode of oscillation and the dipole modes of oscillation.  Octupole modes will be present in the same frequency range but typically at low signal-to-noise ratio.  We deliberately excluded the quadrupole modes to keep the parameter space we explore to an absolute minimum.
\par 
We defined the frequency of the radial mode as a free parameter, $\nu_{n,0}$.  We determined the frequencies of the dipole modes using the roots of the asymptotic expression of \cite{2012A&A...540A.143M} (see also \citealt{2015arXiv150906193M}):
\begin{equation}
\frac{\pi \left( \nu - \nu_{p,1} \right)}{\Delta \nu} = \arctan \left(q \tan \left( \frac{\pi}{\Delta \Pi_{1} \nu} - \epsilon_{g} \right) \right),
\end{equation}
where $\nu$ is the independent variable, $\nu_{p,1}$ is the frequency of the a nominal dipole p mode, $\Delta \nu$ is the large frequency spacing, $q$ is the magnitude of the coupling which is a measure of the p mode and g mode phases, $\Delta \Pi_{1}$ the \edit{asymptotic} dipole mode period spacing, and $\epsilon_{g}$ is a phase term.  We solved for the roots using a sparse grid and interpolation between points.
\par 
Given the frequencies of a set of dipole mixed modes, we added in the effects of rotation that split the dipole modes into a rotational triplet.  The amount of rotational splitting is dependent on the way in which the mode is sensitive to the interior of the star and the stars rotation profile.  Here we used the asymptotic work of  \cite{2015A&A...580A..96D} for a description of the rotation.  We defined the function $\zeta$, which is the mode mixing function,
\begin{equation}
\zeta (\nu) = \left[ 1 + \frac{\alpha(\nu)}{q \beta(\nu)} \right]^{-1},
\end{equation}
where 
\begin{equation}
\alpha (\nu) = \cos^{2} \left( \pi \left( \frac{1}{\nu \, \Delta \Pi_{1}} - \epsilon_{g} \right) \right) \, \nu^{2} \, \Delta \Pi_{1}, 
\end{equation}
and
\begin{equation}
\beta (\nu) = \Delta \nu \, \cos^{2} \left( \pi \left( \frac{\nu - \nu_{p,1}}{\Delta \nu}\right) \right).
\end{equation}

We then calculated the rotational splitting as:
\begin{equation}
\frac{\delta \nu_{nlm}}{m} = \left( \frac{\delta \nu_{g}}{2} - \delta \nu_{p} \right) \zeta(\nu_{nl}) + \delta \nu_{p},
\end{equation}
where $\delta \nu_{g}$ is a splitting parameter for a g-like mode with high inertia, and $\delta \nu_{p}$ is the rotational splitting of a nominal p mode.  The g-like mode splitting is divided 2 to account for the so called Ledoux parameter that accounts for splitting due to the Coriolis force.
\par 
To estimate mode line widths ($\Gamma$) and amplitudes ($A$) we used the mixing function ($\zeta(\nu_{nlm})$).  From \cite{2014ApJ...781L..29B}, we state that the ratio of the inertia of a radial and dipole mode can be approximated using:
\begin{equation}
\frac{I_{n,1}}{I_{n,0}} \simeq 1.5 \, \frac{A_{n,0}}{A_{n,1}} \sqrt{\frac{\Gamma_{n,0}}{\Gamma_{n,1}}},
\label{eq::inert}
\end{equation} 
where the subscripts are labels identifying modes and labelling the mode degree.  Mode amplitude is related to mode height ($H$) as:
\begin{equation}
A^{2} = \frac{\pi}{2} \, \Gamma \, H.  
\end{equation}
We defined the mode linewidth of the dipole mode using $\zeta$ as:
\begin{equation}
\Gamma_{n,1} = \Gamma_{n,0} (1 - \zeta(\nu_{nlm})),
\label{eq::gamma}
\end{equation}
i.e., that mode linewidth for a nominal dipole p mode is the same as the linewidth of the radial mode, and that the line widths of the g-dominated modes are very small.  Modes of oscillation that are unresolved, i.e., those that have very small line widths, are not well modelled in a power spectrum that has finite frequency resolution.  A correct model would convolve the mode shape with the sinc squared response of the power spectrum.  In order to remain computationally efficient, we did not apply this convolution but instead limited the mode linewidth to the frequency resolution or above. 
\par
We then calculated the mode amplitude by the substitution equation \ref{eq::gamma} into \ref{eq::inert}.
\par 
Our model for the signal-to-noise ratio spectrum in our defined range in frequency was then the sum of all the radial and dipole modes plus some flat background term.  Each mode was represented by a Lorentzian profile, to represent a damped harmonic oscillator, giving the model as
\begin{equation}
\begin{aligned}
M(\nu) = W + & \sum_{n}\sum_{l}\sum_{m} \\
& \frac{\varepsilon(\theta, l, m) \, H_{nlm}}{1 + 4 / \Gamma_{nlm}^{2} \left(\nu - \nu_{nlm} - \delta\nu_{nlm} \right)},
\end{aligned}
\end{equation}  
where $W$ is the flat noise background and $\varepsilon(\theta, l, m)$ is a function that accounts for the visibility of the $m$ components given the angle of inclination of the rotation of the star ($\theta$, see \cite{2003ApJ...589.1009G}).
\par 
We assessed the probability of observing the data ($D$) given the model as the negative log likelihood ($S$) using the standard likelihood function \citep{1990ApJ...364..699A}:
\begin{equation}
S = - \ln L = \sum_{i} \ln M + \frac{D}{M},
\end{equation}
where the sum over $i$ represents the sum of each frequency bin in the power spectrum.\\
We estimated the parameters of the model by performing a fit using a parallel tempered affine-invariant Monte Carlo Markov Chain (MCMC) ensemble sampler \citep{2013PASP..125..306F}.  This algorithm was capable of exploring the complex likelihood-parameter space and was able to converge to stable solutions given enough time (that is, given sufficient walkers and iterations).  The algorithm allowed us to estimate the marginalised posterior probability densities for all parameters.  Formally, each parameter must have some prior probability distribution.  These priors were selected based on our a priori knowledge of the characteristics of red giants (i.e., see \cite{2014A&A...572L...5M} and references there in) or were set as broad uninformative or uniform priors. 

\section{Results}

\begin{figure}
\includegraphics[width=\linewidth]{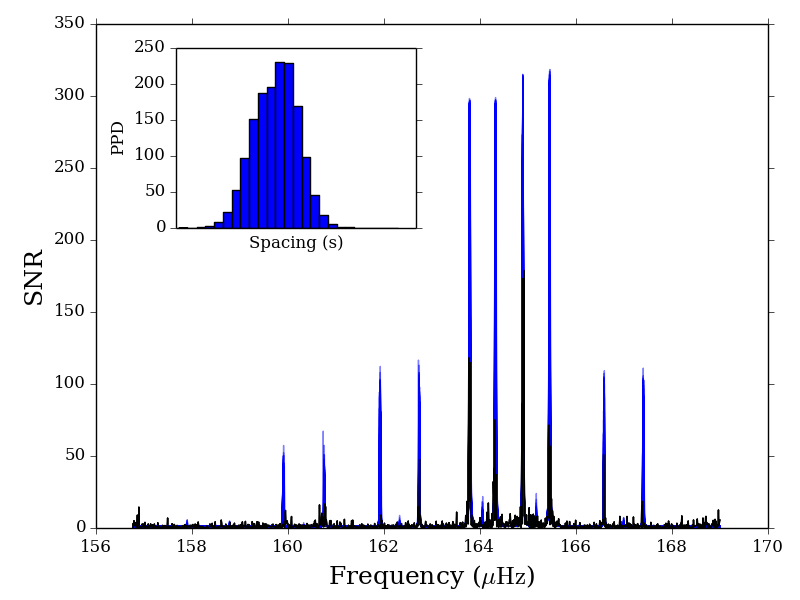} 
\caption{Fit of the model to the signal-to-noise ratio spectrum.  The data are shown in black.  In blue we show models drawn from every 1000th step in the Markov Chains to provide a representation of how well the model fits the data.  The radial mode at $158.0 \rm \mu Hz$ has been removed by fitting and then diving the power spectrum through.  Inset posterior probability distribution for $\Delta \Pi_{1}$ in seconds.}
\label{fig::fit1}
\end{figure}

Figure \ref{fig::fit1} shows the power spectrum and a selection of models that fit the data.  The fitted models clearly reproduce the mode structure in terms of mode frequencies and mode line widths.  The structure in terms of mode amplitude (or height) is less clear to the eye.  The selection of models drawn include a number of models that would appear to overestimate the amplitude of the modes.  In fact, the mean values of mode amplitude or height are much closer to values that are consistent with the data but the tail of the distribution confuses the eye.
\par 
Table \ref{tab::comp} displays the summary statistics of the posterior probability distributions for the parameters that we have fitted.  It is clear that we obtain high precision estimates of the period spacing and the rotational splitting of the g-dominated modes.  Furthermore, the coupling term is well enough constrained to provide insight on the internal structure of the star.
\par  

\edit{
For this study we have examined a low-luminosity red giant.  In a number of ways this presents a best case scenario.   For high-luminosity stars with low $\numax$ we detect only a small number of modes, not the rich spectrum we see in KIC 12008916. As a result, for high-luminosity giants it may not be possible to estimate the g-mode period spacing at all.  Other challenges, such as disentangling period spacing signature from rotational splitting in red clump stars, can be over come with the statistical description of the model and parameters given above.
\par   }

For the purpose of reconstructing the history of the Milky Way, the key derived asteroseismic quantities for red giant stars are the stellar radius (and hence distance), mass, and age.  Excellent constraint on the radius can be gained by considering the radial modes and the dipole nominal p-mode frequencies.  The highest levels of precision and accuracy are believed to be obtained when comparing observed individual frequencies to stellar models. When considering the frequencies of individual modes but care must be exercised to account got the line-of-sight radial velocity contribution to the frequency \citep{2014MNRAS.445L..94D}.  While an ensemble study of many red giant stars is yet to be performed using individual frequencies, for main sequence stars the precision on radius is of order a few percent, mass around $5 \%$, and for age $14 \%$ \citep{2015MNRAS.452.2127S}.

\section{Precision for different space missions}
Table \ref{tab::comp} shows parameters determined from varying length of time series representing different space missions.  Clearly the precision on the measurements increases with increasing length of observation.  As expected, returns diminish as the length of the time series becomes very long (see \cite{2012A&A...544A..90H} for more details on the global properties).  
\begin{table*}
\begin{tabular}{|c|c|c|c|c|c|c|}
\hline 
Parameter & 1335 Days & 730 Days & 351 days & 150 Days & 70 Days & 27 Days \\ 
 &  {\it Kepler} & PLATO & TESS & CoRoT & TESS, or K2 & TESS\\ 
\hline
\hline
$T_{\rm eff}$ (K) & \multicolumn{6}{c|}{$4830 \pm 100$}\\
$\rm [Fe/H]$ & \multicolumn{6}{c|}{$0.05$}\\
\hline
\hline 
$\Delta \nu (\rm \mu \, Hz)$ & $12.89\pm0.01$ & $12.88\pm0.01$ & $12.89\pm0.01$ & $12.90\pm0.02$ & $12.88\pm0.02$ & $12.98\pm0.2$ \\
$\nu_{\rm max} (\rm \mu Hz)$ & $160.9\pm0.5$ & $160.3\pm0.5$ & $160.7\pm0.6$ & $160.6\pm1.0$ & $160.4\pm1.5$ & $161\pm2$ \\
Radius $(R_{\odot})$ & $5.26\pm0.10$ & $5.25\pm0.10$ & $5.25\pm0.1$ & $5.24\pm0.11$ & $5.25\pm0.11$ & $5.2\pm0.2$\\
Mass $(M_{\odot})$ & $1.31\pm0.07$ & $1.30\pm0.07$ & $1.31\pm0.07$ & $1.30\pm0.07$ & $1.31\pm0.08$ & $1.3\pm0.1$\\
\hline 
\hline
$\Delta \Pi_{1}$ (s) & $80.450 \pm 0.002$ & $80.452\pm0.008$ & $80.454\pm0.04$ & $80.46\pm0.25$ & $79.3\pm1.4$ & - \\ 
$q$ & $0.145\pm0.001$ & $0.141\pm0.002$ & $0.138\pm0.003$ & $0.15\pm0.01$ & $0.14\pm0.02$ & - \\ 
$\epsilon_{g}$ & $0.008\pm0.003$ & $0.008\pm0.007$ & $0.0\pm0.1$ & $0.0\pm0.3$ & $0.3\pm1.1$ & - \\ 
$\nu_{p,1}$ ($\rm \mu \, Hz$) & $164.577\pm0.007$ & $164.56\pm0.01$ & $164.57\pm0.02$ & $164.61\pm0.05$ & $163.9\pm0.4$ & - \\ 
$\delta \nu_{g}$ ($\rm \mu \, Hz$) & $0.886\pm0.005$ & $0.87\pm0.01$ & $0.87\pm0.01$ & $0.83\pm0.03$ & $0.88\pm0.1$ & - \\
$\delta \nu_{p}$ $^{\star}$ ($\rm \mu \, Hz$) & $0.002 \pm 0.002$ & $0.006\pm0.008$ & $0.02\pm0.01$ & $0.07\pm 0.04$ & $0.3\pm0.1$ & - \\
$\theta$ ($^{\circ}$) & $88\pm2$ &  $83\pm3$ & $80\pm4$ & $81\pm5$ & $45\pm14$ & - \\ 
\hline 
\end{tabular}
\caption{Parameters obtained by fitting the model to the power spectrum of the {\it Kepler} data set of KIC 12008916 for different length of time series. Each length of data set approximately corresponds to data lengths that may be provided by different space telescopes. Masses and radii are taken from the standard scaling relations (Eq \ref{eq::radius} and \ref{eq::mass}).  $^{\star}$ Note that the posterior probability distribution for the rotational splitting of the envelope is not well described by the summary statistics provided here.} 
\label{tab::comp}
\end{table*}

While radius is a parameter of interest, particularly if estimating distance, it is mass (or at least initial mass) that is the fundamental parameter.  For mass, we can see that reasonable uncertainties of around $10\%$ or better are achievable \edit{by using scaling relations with even} the shortest data sets.  This is important because K2 and TESS (and PLATO in its step-and-stare phase) will cover large areas of sky.  This coverage, and the asteroseismic precision, will allow us to build a detailed picture of stellar mass (and hence age) as a function of location in the galaxy.   

\section{Inferring precise stellar properties}
\begin{figure}
\includegraphics[width=\linewidth]{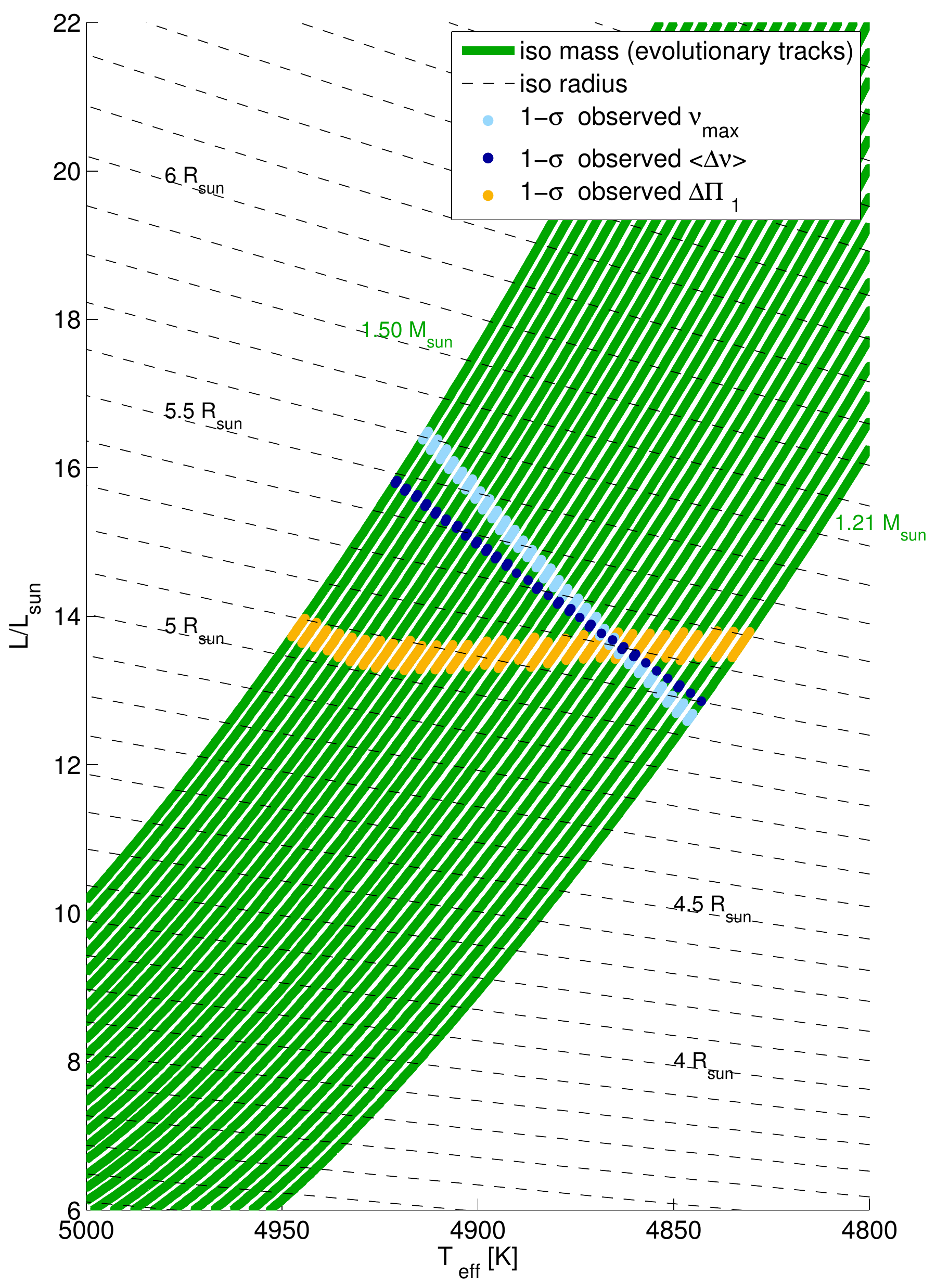} 
\caption{HR diagram representing stellar properties satisfying the combination of predictions from stellar evolutionary tracks ($M=1.21-1.50$  M$_\odot$ in steps of 0.01 M$_\odot$, solid green lines), and the asteroseismic constraints available for KIC 12008916. Dashed black lines represent lines of constant radius (in steps of 0.1 R$_\odot$). As evinced from the plot such a combination of allows an extremely precise - yet model dependent - determination of stellar properties.}
\label{fig::3way}
\end{figure}

As an example of how the seismic indices introduced in Sec. \ref{sec::defs} may be used to infer precise stellar properties, we consider as observational constraints $\dnu$, $\numax$, and $\Delta \Pi_{1}$ as determined from a 150-d time series (see Table \ref{tab::comp}).
To determine stellar properties, we compare such observational constraints with predictions from a set of stellar evolution models. We have used the MESA code \citep{Paxton2011} to compute  stellar evolutionary tracks of solar metallicity with mass from 1.21 M$_\odot$ to 1.50 M$_\odot$ in steps of 0.01 M$_\odot$, adopting the same micro and macro physics prescriptions as in \citet{Bossini2015}.

As evinced from  Fig. \ref{fig::3way},  the combination of observational constraints on $\dnu$,$\numax$, and $\Delta \Pi_{1}$ is in principle able to set exceptionally tight limits on global stellar properties such as mass, radius and luminosity (hence distance).  It is however crucial to  keep in mind that our ability to make use of these constraints may be limited by  uncertainties related to e.g. metallicity and, crucially, by known shortcomings of current stellar models \citep[e.g. see][for a recent review]{Cassisi2014}. We shall discuss in what follows how some of these uncertainties affect the inferred stellar properties. 

\section{From precise to accurate stellar properties}
Seismic data analysis and interpretation techniques have undergone a rapid and considerable development in the last few years. However, they still suffer from limitations, e.g.:
\begin{itemize}
\item determination of individual oscillations mode parameters has been carried out for a limited set of Sun-like stars, and for only a handful of red giants;
\item stellar mass and radius estimates in most cases are based on approximated scaling relations of average seismic properties ($\dnu$ and $\numax$), under-utilising the information content of oscillation modes; and:
\item systematic uncertainties on the inferred stellar properties due to limitations of current stellar models have not yet been quantified. This is crucial for age estimates, which are inherently model dependent.
\end{itemize}


Here, we mention some of the key sources of uncertainty on current estimates of stellar age and how, in some cases, we hope to make progress. We focus on evolved stars, and refer to \citet{Lebreton2014}, and references therein, for the case of main-sequence stars.


Stellar mass is a particularly valuable constraint in the case of giants, since for these stars age is primarily a function of mass. 
The age of low-mass red-giant stars is largely determined by the time spent on the main sequence, hence by the initial mass of the red giant's progenitor ($\tau_{\rm MS} \propto M/L(M) \propto M_{\rm ini}^{-(\nu-1)}$ with  $\nu= 3-5$, e.g. see \citealt{Kippenhahn2012}).

\begin{figure*}
\begin{center}
\includegraphics[width=.48\linewidth]{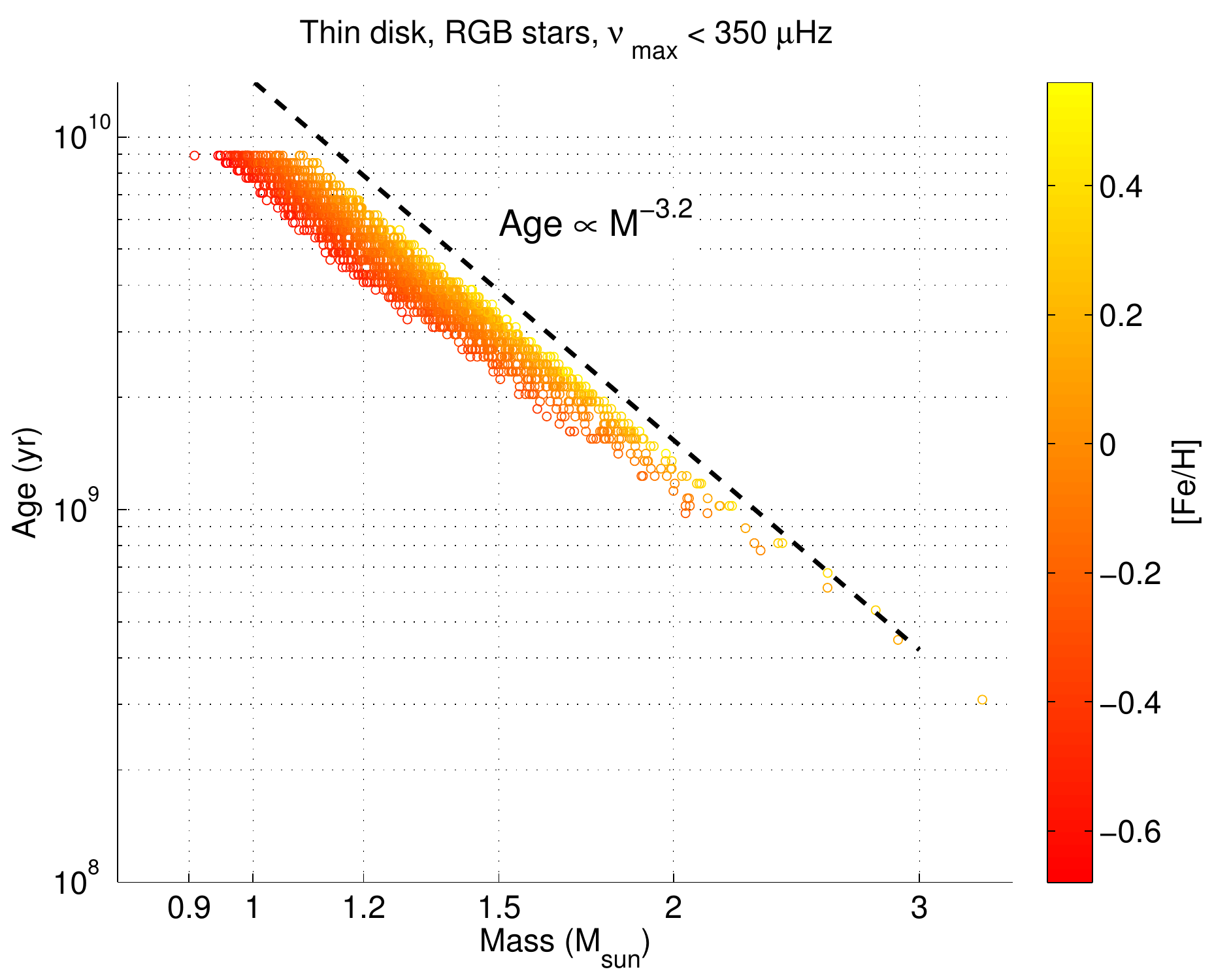} 
\includegraphics[width=.48\linewidth]{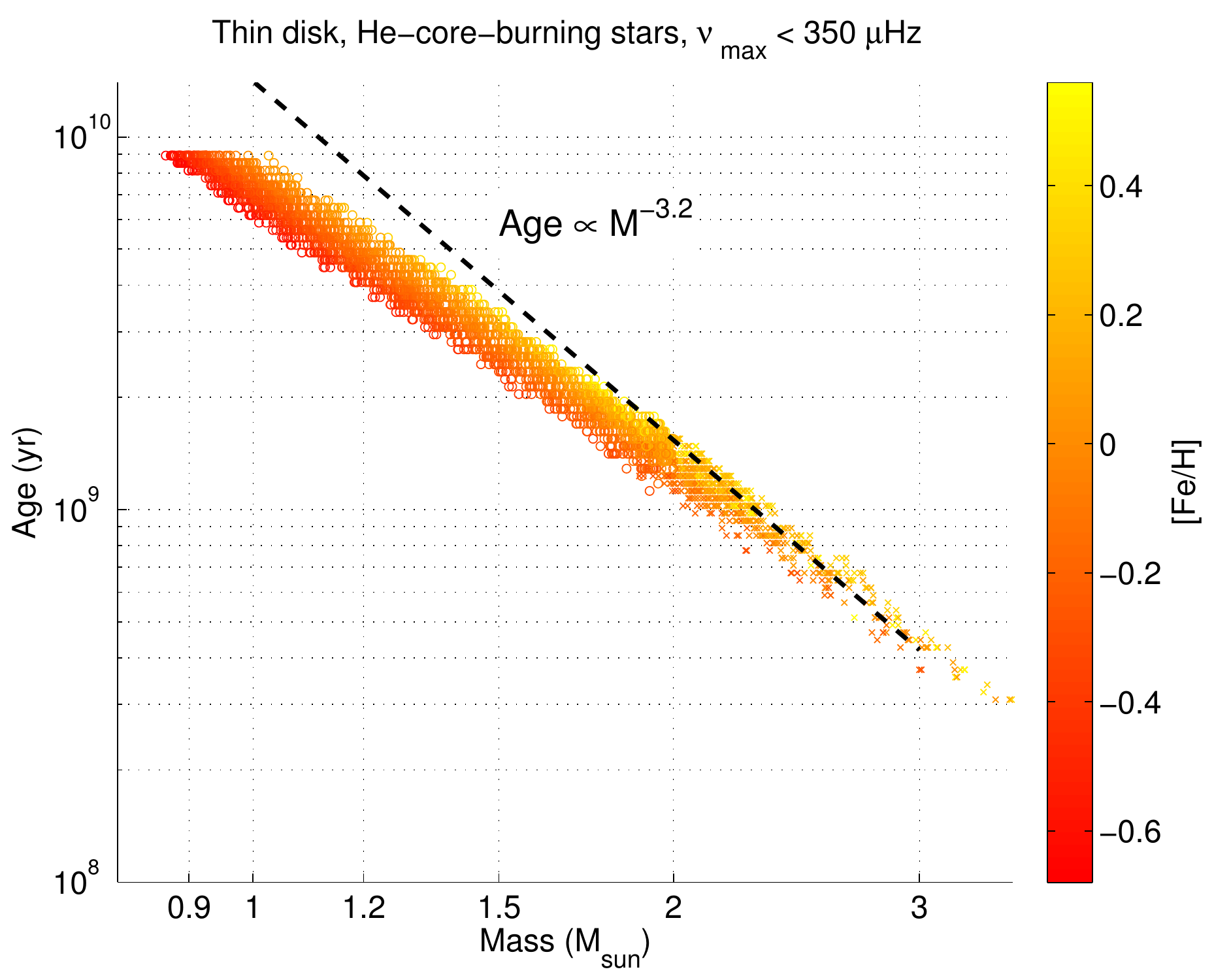} 
\end{center}
\caption{Age-mass-metallicity relation for  red giants in a TRILEGAL synthetic population representative of thin-disk stars observed by {\it Kepler}. While on the RGB (left panel) the age-mass relation follows the expected simple trend,  He-core-burning stars (right panel)  deviate from that relation due to mass loss occurring (following the prescription by \citealt{Reimers1975a} ) near the RGB tip.  Mass loss is expected to affect significantly low-mass stars only.}
\label{fig::age}
\end{figure*}

\edit{
Knowledge of the star's metallicity is also key to determining the age.  Based on predictions from stellar evolutionary tracks,  if one were to consider the stellar mass as known, an uncertainty of 0.1 dex in [Fe/H] would lead to a $~10\%$ uncertainty on the age of a red-giant star (as can be qualitatively inferred also from Fig. \ref{fig::age}).
However, in practice,  constraints on the  chemical composition and mass will be coupled to additional constraints (e.g. radius, $T_{\rm eff}$, $\Delta \Pi_{1}$, luminosity). In reality, chemical composition and mass are constrained by both observables and the requirement that matching models of stellar evolution must satisfy the equations of stellar structure. This leads to a much improved precision on the inferred properties, including age, albeit at the expense of an increased model dependence. More details will be presented in Rodrigues et al, in preparation.}

Given the mass range typical of the observed solar-like oscillating giants (1--3 M$_\odot$),  we can probe $\sim ~1.5$  orders of magnitude in age. 
Figure \ref{fig::age} demonstrates the age-mass relation of giant stars predicted by stellar models. The synthetic population shown in the figure has been computed with the \mbox{TRILEGAL} code \citep{Girardi05}, and is representative of thin-disk star population as observed by the nominal {\it Kepler} mission.

What is challenging, however, is that if we wish to determine ages to 30\% or better, then we need to be able to the determine masses with an accuracy better than 10\%. 
Testing the accuracy of the asteroseismic mass scale to 10\% or better is very much ``work in progress''. 
Comparisons against accurate and independent mass determinations are, however,  limited to stars in binary systems and, most notably, stars in clusters (for a review see, e.g., Brogaard et al., this volume). 

An example of possible systematic biases concerning the mass determination 
are departures from a simple scaling of $\dnu$  with the square root of the stellar mean density  \citep[see e.g.][]{White2011,Miglio2012, Miglio2013, Belkacem2013}. 
Suggested corrections to the $\dnu$\ scaling are likely to depend (to a level of few percent) on the the stellar structure itself. Moreover, the average $\dnu$\ is known to be affected (to a level of $\sim 1\%$ in the Sun) by our inaccurate modelling of near-surface layers. In most cases the main effect of using model-predicted $\dnu$ is a reduction $\approx 10\%$ or less of the mass estimate for RGB stars based on Eq. \ref{eq::mass}.
A thorough description of the $\dnu$ corrections, their limitations and their dependences on stellar properties, will be presented in Rodrigues et al., in preparation. 

A way forward would be to determine the star's mean density by using the full set of observed acoustic modes, not just their average frequency spacing. 
This approach was carried out in at least two RGB stars \citep{Huber2013, Lillo-Box2014}, and led to determination of the stellar mean density which is $\sim 5-6\%$ higher than derived from assuming scaling relations, and with a much improved precision of  $\sim 1.4\%$.

While a relatively simple mass-age relation is expected for RGB stars (Fig. \ref{fig::age}, left panel),  the situation for red-clump (RC) or early asymptotic-giant-branch (AGB) stars is  different (Fig. \ref{fig::age}, right panel). 
 If stars undergo a significant mass loss near the tip of the RGB, then the mass-age relation is not unique (for a given composition and input physics), since the mass observed at the RC or early-AGB stage may differ from the initial one. 
From a closer inspection of Fig. \ref{fig::age}, it is worth noticing  that for stars with $M < 1.5$ M$_\odot$ the age-mass relation bifurcates due to the significant  mass loss ($\sim 0.1-0.2$ M$_\odot$) experienced by low-mass stars near the tip of the RGB (if one adopts the mass-loss prescription by \citealt{Reimers1975a}). Consequently, RC stars are younger than stars on the  RGB with the same actual mass (and metallicitiy). We can, however, remove this degeneracy in the age-mass relation thanks to additional seismic constraint, particularly from the g-mode period spacing $\Delta \Pi_{1}$,
which  allows a clear distinction to be made
between RGB and RC stars \citep{Bedding2011}, and early-AGB stars \citep{Montalban2013a}.
Knowledge of the efficiency of mass loss is however still needed to determine accurate ages of RC stars (see e.g. \citealt{Miglio2012} and \citealt{Handberg2015} for a discussion on seismic constraints on mass loss efficiency).

When discussing systematic uncertainties on age estimates, it is worth recalling that uncertainties on the input physics may affect main-sequence lifetimes, hence the age of red giants. A thorough comparison of age predictions from various stellar evolution codes, and with different assumptions concerning the input physics, will be presented in Miglio et al. in preparation (see also \texttt{http://www.asterostep.eu/Projects.html}).

Additional seismic diagnostics are still to be fully utilised and their dependence on stellar properties understood. 
The use of the dipole g-mode period spacing to infer mass, although very promising (see Fig. \ref{fig::3way}), is in its infancy.  Studies are progressing that are testing the robustness of the use of the $\Delta \Pi_{1}$ parameter, testing for any bias between observed and model values, and the dependence on mixing processes in prior evolutionary states \citep{Lagarde2015}.
Seismic signatures of sharp-structure variations can potentially lead to estimates of the envelope He abundance \citep[see ][]{Broomhall2014}, or to detailed constraints on  near-core regions \citep{Montalban2013, Bossini2015, Constantino2015, Cunha2015}. Promising indicators of global stellar properties include the small separation between radial and quadrupole modes  \citep{Montalban2012} and the properties of mixed modes and coupling term which may lead to additional indirect constraints on the stellar mass \citep[see e.g.][]{Benomar2013}. 

\section{Summary and outlook}

Current space missions ({\it Kepler}, CoRoT, K2) are providing, and future missions (K2, TESS, PLATO) will provide, a wealth of observational photometric data that we are learning to efficiently and precisely interpret for the purposes of asteroseismology.  Here we have demonstrated methods for analysing the power spectra of a low-luminosity red giant and comparison of the resulting observational parameters with stellar models.  When able to leverage the global properties ($\Delta \nu$, $\nu_{\rm max}$) and the g-mode property ($\Delta \Pi_{1}$) we are able to derive high precision masses, and hence ages.  We have shown that for data sets of varying temporal length, representing the different space missions, we are able to estimate all three critical seismic parameters for all but the shortest data sets.  

However, we are still faced with a number of challenges that are focused around tests of the accuracy of stellar models that determine the properties of stellar populations can be determined.  Robust predictions from stellar models are key to determining accurate stellar properties such as mass, radius, surface gravity and, crucially, age. A critical appraisal of how numerical and systematic uncertainties in model predictions impact the inferred stellar properties (in particular ages) is needed. In favourable cases (such as binary systems, clusters) stellar models will be tested against the seismic measurements and reduce (some of) the systematic uncertainties in the age determination related to, for example, near-core extra mixing during the main sequence, and mass loss on the red-giant branch. Given the additional constraints (stringent priors on age, chemical composition) stars in clusters  and binary systems represent the prime targets for testing models.

\section*{Acknowledgements}
We acknowledge funding from the Wilhelm and Else Heraeus Foundation and the support of the UK Science and Technology
Facilities Council (STFC).  

\bibliographystyle{an}
\bibliography{refs}

\begin{thebibliography}{51}
\expandafter\ifx\csname natexlab\endcsname\relax\def\natexlab#1{#1}\fi

\bibitem[{{Anderson} {et~al.}(1990){Anderson}, {Duvall}, \&
  {Jefferies}}]{1990ApJ...364..699A}
{Anderson}, E.~R., {Duvall}, Jr., T.~L., \& {Jefferies}, S.~M. 1990, \apj, 364,
  699

\bibitem[{{Baglin} {et~al.}(2006){Baglin}, {Michel}, {Auvergne}, \& {COROT
  Team}}]{2006ESASP.624E..34B}
{Baglin}, A., {Michel}, E., {Auvergne}, M., \& {COROT Team}. 2006, in ESA
  Special Publication, Vol. 624, Proceedings of SOHO 18/GONG 2006/HELAS I,
  Beyond the spherical Sun, 34

\bibitem[{{Bedding} {et~al.}(2011){Bedding}, {Mosser}, {Huber},
  {Montalb{\'a}n}, {Beck}, {Christensen-Dalsgaard}, {Elsworth},
  {Garc{\'{\i}}a}, {Miglio}, {Stello}, {White}, {De Ridder}, {Hekker}, {Aerts},
  {Barban}, {Belkacem}, {Broomhall}, {Brown}, {Buzasi}, {Carrier}, {Chaplin},
  {di Mauro}, {Dupret}, {Frandsen}, {Gilliland}, {Goupil}, {Jenkins},
  {Kallinger}, {Kawaler}, {Kjeldsen}, {Mathur}, {Noels}, {Aguirre}, \&
  {Ventura}}]{Bedding2011}
{Bedding}, T.~R., {Mosser}, B., {Huber}, D., {et~al.} 2011, \nat, 471, 608

\bibitem[{{Belkacem} {et~al.}(2013){Belkacem}, {Samadi}, {Mosser}, {Goupil}, \&
  {Ludwig}}]{Belkacem2013}
{Belkacem}, K., {Samadi}, R., {Mosser}, B., {Goupil}, M.-J., \& {Ludwig}, H.-G.
  2013, in Astronomical Society of the Pacific Conference Series, Vol. 479,
  Progress in Physics of the Sun and Stars: A New Era in Helio- and
  Asteroseismology, ed. H.~{Shibahashi} \& A.~E. {Lynas-Gray}, 61

\bibitem[{{Benomar} {et~al.}(2013){Benomar}, {Bedding}, {Mosser}, {Stello},
  {Belkacem}, {Garcia}, {White}, {Kuehn}, {Deheuvels}, \&
  {Christensen-Dalsgaard}}]{Benomar2013}
{Benomar}, O., {Bedding}, T.~R., {Mosser}, B., {et~al.} 2013, \apj, 767, 158

\bibitem[{{Benomar} {et~al.}(2014){Benomar}, {Belkacem}, {Bedding}, {Stello},
  {Di Mauro}, {Ventura}, {Mosser}, {Goupil}, {Samadi}, \&
  {Garcia}}]{2014ApJ...781L..29B}
{Benomar}, O., {Belkacem}, K., {Bedding}, T.~R., {et~al.} 2014, \apjl, 781, L29

\bibitem[{{Borucki} {et~al.}(2010){Borucki}, {Koch}, {Basri}, {Batalha},
  {Brown}, {Caldwell}, {Caldwell}, {Christensen-Dalsgaard}, {Cochran},
  {DeVore}, {Dunham}, {Dupree}, {Gautier}, {Geary}, {Gilliland}, {Gould},
  {Howell}, {Jenkins}, {Kondo}, {Latham}, {Marcy}, {Meibom}, {Kjeldsen},
  {Lissauer}, {Monet}, {Morrison}, {Sasselov}, {Tarter}, {Boss}, {Brownlee},
  {Owen}, {Buzasi}, {Charbonneau}, {Doyle}, {Fortney}, {Ford}, {Holman},
  {Seager}, {Steffen}, {Welsh}, {Rowe}, {Anderson}, {Buchhave}, {Ciardi},
  {Walkowicz}, {Sherry}, {Horch}, {Isaacson}, {Everett}, {Fischer}, {Torres},
  {Johnson}, {Endl}, {MacQueen}, {Bryson}, {Dotson}, {Haas}, {Kolodziejczak},
  {Van Cleve}, {Chandrasekaran}, {Twicken}, {Quintana}, {Clarke}, {Allen},
  {Li}, {Wu}, {Tenenbaum}, {Verner}, {Bruhweiler}, {Barnes}, \&
  {Prsa}}]{2010Sci...327..977B}
{Borucki}, W.~J., {Koch}, D., {Basri}, G., {et~al.} 2010, Science, 327, 977

\bibitem[{{Bossini} {et~al.}(2015){Bossini}, {Miglio}, {Salaris},
  {Pietrinferni}, {Montalb{\'a}n}, {Bressan}, {Noels}, {Cassisi}, {Girardi}, \&
  {Marigo}}]{Bossini2015}
{Bossini}, D., {Miglio}, A., {Salaris}, M., {et~al.} 2015, \mnras, 453, 2290

\bibitem[{{Broomhall} {et~al.}(2014){Broomhall}, {Miglio}, {Montalb{\'a}n},
  {Eggenberger}, {Chaplin}, {Elsworth}, {Scuflaire}, {Ventura}, \&
  {Verner}}]{Broomhall2014}
{Broomhall}, A.-M., {Miglio}, A., {Montalb{\'a}n}, J., {et~al.} 2014, \mnras,
  440, 1828

\bibitem[{{Cassisi}(2014)}]{Cassisi2014}
{Cassisi}, S. 2014, in EAS Publications Series, Vol.~65, EAS Publications
  Series, 17--74

\bibitem[{{Chaplin} {et~al.}(2015){Chaplin}, {Lund}, {Handberg}, {Basu},
  {Buchhave}, {Campante}, {Davies}, {Huber}, {Latham}, {Latham}, {Serenelli},
  {Antia}, {Appourchaux}, {Ball}, {Benomar}, {Casagrande},
  {Christensen-Dalsgaard}, {Coelho}, {Creevey}, {Elsworth}, {Garc}, {Gaulme},
  {Hekker}, {Kallinger}, {Karoff}, {Kawaler}, {Kjeldsen}, {Lundkvist},
  {Marcadon}, {Mathur}, {Miglio}, {Mosser}, {R}, {Roxburgh}, {Silva Aguirre},
  {Stello}, {Verma}, {White}, {Bedding}, {Barclay}, {Buzasi}, {Deheuvels},
  {Gizon}, {Houdek}, {Howell}, {Salabert}, \&
  {Soderblom}}]{2015arXiv150701827C}
{Chaplin}, W.~J., {Lund}, M.~N., {Handberg}, R., {et~al.} 2015, ArXiv e-prints

\bibitem[{{Chaplin} \& {Miglio}(2013)}]{2013ARA&A..51..353C}
{Chaplin}, W.~J. \& {Miglio}, A. 2013, \araa, 51, 353

\bibitem[{{Constantino} {et~al.}(2015){Constantino}, {Campbell},
  {Christensen-Dalsgaard}, {Lattanzio}, \& {Stello}}]{Constantino2015}
{Constantino}, T., {Campbell}, S.~W., {Christensen-Dalsgaard}, J., {Lattanzio},
  J.~C., \& {Stello}, D. 2015, \mnras, 452, 123

\bibitem[{{Corsaro} {et~al.}(2015){Corsaro}, {De Ridder}, \&
  {Garc{\'{\i}}a}}]{2015A&A...579A..83C}
{Corsaro}, E., {De Ridder}, J., \& {Garc{\'{\i}}a}, R.~A. 2015, \aap, 579, A83

\bibitem[{{Cunha} {et~al.}(2015){Cunha}, {Stello}, {Avelino},
  {Christensen-Dalsgaard}, \& {Townsend}}]{Cunha2015}
{Cunha}, M.~S., {Stello}, D., {Avelino}, P.~P., {Christensen-Dalsgaard}, J., \&
  {Townsend}, R.~H.~D. 2015, \apj, 805, 127

\bibitem[{{Davies} {et~al.}(2015){Davies}, {Chaplin}, {Farr}, {Garc{\'{\i}}a},
  {Lund}, {Mathis}, {Metcalfe}, {Appourchaux}, {Basu}, {Benomar}, {Campante},
  {Ceillier}, {Elsworth}, {Handberg}, {Salabert}, \&
  {Stello}}]{2015MNRAS.446.2959D}
{Davies}, G.~R., {Chaplin}, W.~J., {Farr}, W.~M., {et~al.} 2015, \mnras, 446,
  2959

\bibitem[{{Davies} {et~al.}(2014){Davies}, {Handberg}, {Miglio}, {Campante},
  {Chaplin}, \& {Elsworth}}]{2014MNRAS.445L..94D}
{Davies}, G.~R., {Handberg}, R., {Miglio}, A., {et~al.} 2014, \mnras, 445, L94

\bibitem[{{Deheuvels} {et~al.}(2015){Deheuvels}, {Ballot}, {Beck}, {Mosser},
  {{\O}stensen}, {Garc{\'{\i}}a}, \& {Goupil}}]{2015A&A...580A..96D}
{Deheuvels}, S., {Ballot}, J., {Beck}, P.~G., {et~al.} 2015, \aap, 580, A96

\bibitem[{{Foreman-Mackey} {et~al.}(2013){Foreman-Mackey}, {Hogg}, {Lang}, \&
  {Goodman}}]{2013PASP..125..306F}
{Foreman-Mackey}, D., {Hogg}, D.~W., {Lang}, D., \& {Goodman}, J. 2013, \pasp,
  125, 306

\bibitem[{{Garc{\'{\i}}a} {et~al.}(2011){Garc{\'{\i}}a}, {Hekker}, {Stello},
  {Guti{\'e}rrez-Soto}, {Handberg}, {Huber}, {Karoff}, {Uytterhoeven},
  {Appourchaux}, {Chaplin}, {Elsworth}, {Mathur}, {Ballot},
  {Christensen-Dalsgaard}, {Gilliland}, {Houdek}, {Jenkins}, {Kjeldsen},
  {McCauliff}, {Metcalfe}, {Middour}, {Molenda-Zakowicz}, {Monteiro}, {Smith},
  \& {Thompson}}]{2011MNRAS.414L...6G}
{Garc{\'{\i}}a}, R.~A., {Hekker}, S., {Stello}, D., {et~al.} 2011, \mnras, 414,
  L6

\bibitem[{Girardi {et~al.}(2005)Girardi, Groenewegen, Hatziminaoglou, \&
  da~Costa}]{Girardi05}
Girardi, L., Groenewegen, M.~A.~T., Hatziminaoglou, E., \& da~Costa, L. 2005,
  \aap, 436, 895

\bibitem[{{Gizon} \& {Solanki}(2003)}]{2003ApJ...589.1009G}
{Gizon}, L. \& {Solanki}, S.~K. 2003, \apj, 589, 1009

\bibitem[{{Handberg} {et~al.}(2015){Handberg}, {Brogaard}, {Miglio},
  {Elsworth}, {Bossini}, \& R.}]{Handberg2015}
{Handberg}, R., {Brogaard}, K.~F., {Miglio}, A., {et~al.} 2015, \mnras,
  submitted

\bibitem[{{Hekker} {et~al.}(2012){Hekker}, {Elsworth}, {Mosser}, {Kallinger},
  {Chaplin}, {De Ridder}, {Garc{\'{\i}}a}, {Stello}, {Clarke}, {Hall}, \&
  {Ibrahim}}]{2012A&A...544A..90H}
{Hekker}, S., {Elsworth}, Y., {Mosser}, B., {et~al.} 2012, \aap, 544, A90

\bibitem[{{Howell} {et~al.}(2014){Howell}, {Sobeck}, {Haas}, {Still},
  {Barclay}, {Mullally}, {Troeltzsch}, {Aigrain}, {Bryson}, {Caldwell},
  {Chaplin}, {Cochran}, {Huber}, {Marcy}, {Miglio}, {Najita}, {Smith},
  {Twicken}, \& {Fortney}}]{2014PASP..126..398H}
{Howell}, S.~B., {Sobeck}, C., {Haas}, M., {et~al.} 2014, \pasp, 126, 398

\bibitem[{{Huber} {et~al.}(2010){Huber}, {Bedding}, {Stello}, {Mosser},
  {Mathur}, {Kallinger}, {Hekker}, {Elsworth}, {Buzasi}, {De Ridder},
  {Gilliland}, {Kjeldsen}, {Chaplin}, {Garc{\'{\i}}a}, {Hale}, {Preston},
  {White}, {Borucki}, {Christensen-Dalsgaard}, {Clarke}, {Jenkins}, \&
  {Koch}}]{2010ApJ...723.1607H}
{Huber}, D., {Bedding}, T.~R., {Stello}, D., {et~al.} 2010, \apj, 723, 1607

\bibitem[{{Huber} {et~al.}(2013){Huber}, {Carter}, {Barbieri}, {Miglio},
  {Deck}, {Fabrycky}, {Montet}, {Buchhave}, {Chaplin}, {Hekker},
  {Montalb{\'a}n}, {Sanchis-Ojeda}, {Basu}, {Bedding}, {Campante},
  {Christensen-Dalsgaard}, {Elsworth}, {Stello}, {Arentoft}, {Ford},
  {Gilliland}, {Handberg}, {Howard}, {Isaacson}, {Johnson}, {Karoff},
  {Kawaler}, {Kjeldsen}, {Latham}, {Lund}, {Lundkvist}, {Marcy}, {Metcalfe},
  {Silva Aguirre}, \& {Winn}}]{Huber2013}
{Huber}, D., {Carter}, J.~A., {Barbieri}, M., {et~al.} 2013, Science, 342, 331

\bibitem[{{Kallinger} {et~al.}(2014){Kallinger}, {De Ridder}, {Hekker},
  {Mathur}, {Mosser}, {Gruberbauer}, {Garc{\'{\i}}a}, {Karoff}, \&
  {Ballot}}]{2014A&A...570A..41K}
{Kallinger}, T., {De Ridder}, J., {Hekker}, S., {et~al.} 2014, \aap, 570, A41

\bibitem[{{Kippenhahn} {et~al.}(2012){Kippenhahn}, {Weigert}, \&
  {Weiss}}]{Kippenhahn2012}
{Kippenhahn}, R., {Weigert}, A., \& {Weiss}, A. 2012, {Stellar Structure and
  Evolution}

\bibitem[{{Lagarde} {et~al.}(2015){Lagarde}, {Bossini}, {Miglio}, {Vrard}, \&
  {Mosser}}]{Lagarde2015}
{Lagarde}, N., {Bossini}, D., {Miglio}, A., {Vrard}, M., \& {Mosser}, B. 2015,
  \mnras, submitted

\bibitem[{{Lebreton} {et~al.}(2014){Lebreton}, {Goupil}, \&
  {Montalb{\'a}n}}]{Lebreton2014}
{Lebreton}, Y., {Goupil}, M.~J., \& {Montalb{\'a}n}, J. 2014, in EAS
  Publications Series, Vol.~65, EAS Publications Series, 177--223

\bibitem[{{Lillo-Box} {et~al.}(2014){Lillo-Box}, {Barrado}, {Moya},
  {Montesinos}, {Montalb{\'a}n}, {Bayo}, {Barbieri}, {R{\'e}gulo}, {Mancini},
  {Bouy}, \& {Henning}}]{Lillo-Box2014}
{Lillo-Box}, J., {Barrado}, D., {Moya}, A., {et~al.} 2014, \aap, 562, A109

\bibitem[{{Meibom} {et~al.}(2015){Meibom}, {Barnes}, {Platais}, {Gilliland},
  {Latham}, \& {Mathieu}}]{2015Natur.517..589M}
{Meibom}, S., {Barnes}, S.~A., {Platais}, I., {et~al.} 2015, \nat, 517, 589

\bibitem[{{Metcalfe} {et~al.}(2015){Metcalfe}, {Creevey}, \&
  {Davies}}]{2015ApJ...811L..37M}
{Metcalfe}, T.~S., {Creevey}, O.~L., \& {Davies}, G.~R. 2015, \apjl, 811, L37

\bibitem[{{Miglio} {et~al.}(2012){Miglio}, {Brogaard}, {Stello}, {Chaplin},
  {D'Antona}, {Montalb{\'a}n}, {Basu}, {Bressan}, {Grundahl}, {Pinsonneault},
  {Serenelli}, {Elsworth}, {Hekker}, {Kallinger}, {Mosser}, {Ventura},
  {Bonanno}, {Noels}, {Silva Aguirre}, {Szabo}, {Li}, {McCauliff}, {Middour},
  \& {Kjeldsen}}]{Miglio2012}
{Miglio}, A., {Brogaard}, K., {Stello}, D., {et~al.} 2012, \mnras, 419, 2077

\bibitem[{{Miglio} {et~al.}(2013{\natexlab{a}}){Miglio}, {Chiappini}, {Morel},
  {Barbieri}, {Chaplin}, {Girardi}, {Montalb{\'a}n}, {Noels}, {Valentini},
  {Mosser}, {Baudin}, {Casagrande}, {Fossati}, {Silva Aguirre}, \&
  {Baglin}}]{Miglio2013}
{Miglio}, A., {Chiappini}, C., {Morel}, T., {et~al.} 2013{\natexlab{a}}, in
  European Physical Journal Web of Conferences, Vol.~43, European Physical
  Journal Web of Conferences, 3004

\bibitem[{{Miglio} {et~al.}(2013{\natexlab{b}}){Miglio}, {Chiappini}, {Morel},
  {Barbieri}, {Chaplin}, {Girardi}, {Montalb{\'a}n}, {Valentini}, {Mosser},
  {Baudin}, {Casagrande}, {Fossati}, {Aguirre}, \& {Baglin}}]{Miglio2013a}
{Miglio}, A., {Chiappini}, C., {Morel}, T., {et~al.} 2013{\natexlab{b}},
  \mnras, 429, 423

\bibitem[{{Montalb{\'a}n} {et~al.}(2013){Montalb{\'a}n}, {Miglio}, {Noels},
  {Dupret}, {Scuflaire}, \& {Ventura}}]{Montalban2013}
{Montalb{\'a}n}, J., {Miglio}, A., {Noels}, A., {et~al.} 2013, \apj, 766, 118

\bibitem[{{Montalb{\'a}n} {et~al.}(2012){Montalb{\'a}n}, {Miglio}, {Noels},
  {Scuflaire}, {Ventura}, \& {D'Antona}}]{Montalban2012}
{Montalb{\'a}n}, J., {Miglio}, A., {Noels}, A., {et~al.} 2012, {Adiabatic
  Solar-Like Oscillations in Red Giant Stars}, ed. A.~{Miglio},
  J.~{Montalb{\'a}n}, \& A.~{Noels}, 23

\bibitem[{{Montalb{\'a}n} \& {Noels}(2013)}]{Montalban2013a}
{Montalb{\'a}n}, J. \& {Noels}, A. 2013, in European Physical Journal Web of
  Conferences, Vol.~43, European Physical Journal Web of Conferences, 3002

\bibitem[{{Mosser} {et~al.}(2014){Mosser}, {Benomar}, {Belkacem}, {Goupil},
  {Lagarde}, {Michel}, {Lebreton}, {Stello}, {Vrard}, {Barban}, {Bedding},
  {Deheuvels}, {Chaplin}, {De Ridder}, {Elsworth}, {Montalban}, {Noels},
  {Ouazzani}, {Samadi}, {White}, \& {Kjeldsen}}]{2014A&A...572L...5M}
{Mosser}, B., {Benomar}, O., {Belkacem}, K., {et~al.} 2014, \aap, 572, L5

\bibitem[{{Mosser} {et~al.}(2012){Mosser}, {Goupil}, {Belkacem}, {Michel},
  {Stello}, {Marques}, {Elsworth}, {Barban}, {Beck}, {Bedding}, {De Ridder},
  {Garc{\'{\i}}a}, {Hekker}, {Kallinger}, {Samadi}, {Stumpe}, {Barclay}, \&
  {Burke}}]{2012A&A...540A.143M}
{Mosser}, B., {Goupil}, M.~J., {Belkacem}, K., {et~al.} 2012, \aap, 540, A143

\bibitem[{{Mosser} {et~al.}(2015){Mosser}, {Vrard}, {Belkacem}, {Deheuvels}, \&
  {Goupil}}]{2015arXiv150906193M}
{Mosser}, B., {Vrard}, M., {Belkacem}, K., {Deheuvels}, S., \& {Goupil}, M.~J.
  2015, ArXiv e-prints

\bibitem[{{Paxton} {et~al.}(2011){Paxton}, {Bildsten}, {Dotter}, {Herwig},
  {Lesaffre}, \& {Timmes}}]{Paxton2011}
{Paxton}, B., {Bildsten}, L., {Dotter}, A., {et~al.} 2011, \apjs, 192, 3

\bibitem[{{Rauer} {et~al.}(2014){Rauer}, {Catala}, {Aerts}, {Appourchaux},
  {Benz}, {Brandeker}, {Christensen-Dalsgaard}, {Deleuil}, {Gizon}, {Goupil},
  {G{\"u}del}, {Janot-Pacheco}, {Mas-Hesse}, {Pagano}, {Piotto}, {Pollacco},
  {Santos}, {Smith}, {Su{\'a}rez}, {Szab{\'o}}, {Udry}, {Adibekyan}, {Alibert},
  {Almenara}, {Amaro-Seoane}, {Eiff}, {Asplund}, {Antonello}, {Barnes},
  {Baudin}, {Belkacem}, {Bergemann}, {Bihain}, {Birch}, {Bonfils}, {Boisse},
  {Bonomo}, {Borsa}, {Brand{\~a}o}, {Brocato}, {Brun}, {Burleigh}, {Burston},
  {Cabrera}, {Cassisi}, {Chaplin}, {Charpinet}, {Chiappini}, {Church},
  {Csizmadia}, {Cunha}, {Damasso}, {Davies}, {Deeg}, {D{\'{\i}}az}, {Dreizler},
  {Dreyer}, {Eggenberger}, {Ehrenreich}, {Eigm{\"u}ller}, {Erikson}, {Farmer},
  {Feltzing}, {de Oliveira Fialho}, {Figueira}, {Forveille}, {Fridlund},
  {Garc{\'{\i}}a}, {Giommi}, {Giuffrida}, {Godolt}, {Gomes da Silva},
  {Granzer}, {Grenfell}, {Grotsch-Noels}, {G{\"u}nther}, {Haswell}, {Hatzes},
  {H{\'e}brard}, {Hekker}, {Helled}, {Heng}, {Jenkins}, {Johansen},
  {Khodachenko}, {Kislyakova}, {Kley}, {Kolb}, {Krivova}, {Kupka}, {Lammer},
  {Lanza}, {Lebreton}, {Magrin}, {Marcos-Arenal}, {Marrese}, {Marques},
  {Martins}, {Mathis}, {Mathur}, {Messina}, {Miglio}, {Montalban}, {Montalto},
  {Monteiro}, {Moradi}, {Moravveji}, {Mordasini}, {Morel}, {Mortier},
  {Nascimbeni}, {Nelson}, {Nielsen}, {Noack}, {Norton}, {Ofir}, {Oshagh},
  {Ouazzani}, {P{\'a}pics}, {Parro}, {Petit}, {Plez}, {Poretti}, {Quirrenbach},
  {Ragazzoni}, {Raimondo}, {Rainer}, {Reese}, {Redmer}, {Reffert},
  {Rojas-Ayala}, {Roxburgh}, {Salmon}, {Santerne}, {Schneider}, {Schou},
  {Schuh}, {Schunker}, {Silva-Valio}, {Silvotti}, {Skillen}, {Snellen}, {Sohl},
  {Sousa}, {Sozzetti}, {Stello}, {Strassmeier}, {{\v S}vanda}, {Szab{\'o}},
  {Tkachenko}, {Valencia}, {Van Grootel}, {Vauclair}, {Ventura}, {Wagner},
  {Walton}, {Weingrill}, {Werner}, {Wheatley}, \&
  {Zwintz}}]{2014ExA....38..249R}
{Rauer}, H., {Catala}, C., {Aerts}, C., {et~al.} 2014, Experimental Astronomy,
  38, 249

\bibitem[{{Reimers}(1975)}]{Reimers1975a}
{Reimers}, D. 1975, M{\'e}moires of the Soci{\'e}t{\'e} Royale des Sciences de
  Li{\`e}ge, 8, 369

\bibitem[{{Ricker} {et~al.}(2014){Ricker}, {Winn}, {Vanderspek}, {Latham},
  {Bakos}, {Bean}, {Berta-Thompson}, {Brown}, {Buchhave}, {Butler}, {Butler},
  {Chaplin}, {Charbonneau}, {Christensen-Dalsgaard}, {Clampin}, {Deming},
  {Doty}, {De Lee}, {Dressing}, {Dunham}, {Endl}, {Fressin}, {Ge}, {Henning},
  {Holman}, {Howard}, {Ida}, {Jenkins}, {Jernigan}, {Johnson}, {Kaltenegger},
  {Kawai}, {Kjeldsen}, {Laughlin}, {Levine}, {Lin}, {Lissauer}, {MacQueen},
  {Marcy}, {McCullough}, {Morton}, {Narita}, {Paegert}, {Palle}, {Pepe},
  {Pepper}, {Quirrenbach}, {Rinehart}, {Sasselov}, {Sato}, {Seager},
  {Sozzetti}, {Stassun}, {Sullivan}, {Szentgyorgyi}, {Torres}, {Udry}, \&
  {Villasenor}}]{2014SPIE.9143E..20R}
{Ricker}, G.~R., {Winn}, J.~N., {Vanderspek}, R., {et~al.} 2014, in Society of
  Photo-Optical Instrumentation Engineers (SPIE) Conference Series, Vol. 9143,
  Society of Photo-Optical Instrumentation Engineers (SPIE) Conference Series,
  20

\bibitem[{{Silva Aguirre} {et~al.}(2015){Silva Aguirre}, {Davies}, {Basu},
  {Christensen-Dalsgaard}, {Creevey}, {Metcalfe}, {Bedding}, {Casagrande},
  {Handberg}, {Lund}, {Nissen}, {Chaplin}, {Huber}, {Serenelli}, {Stello}, {Van
  Eylen}, {Campante}, {Elsworth}, {Gilliland}, {Hekker}, {Karoff}, {Kawaler},
  {Kjeldsen}, \& {Lundkvist}}]{2015MNRAS.452.2127S}
{Silva Aguirre}, V., {Davies}, G.~R., {Basu}, S., {et~al.} 2015, \mnras, 452,
  2127

\bibitem[{{Stello} {et~al.}(2015){Stello}, {Huber}, {Sharma}, {Johnson},
  {Lund}, {Handberg}, {Buzasi}, {Silva Aguirre}, {Chaplin}, {Miglio},
  {Pinsonneault}, {Basu}, {Bedding}, {Bland-Hawthorn}, {Casagrande}, {Davies},
  {Elsworth}, {Garcia}, {Mathur}, {Di Mauro}, {Mosser}, {Schneider},
  {Serenelli}, \& {Valentini}}]{2015ApJ...809L...3S}
{Stello}, D., {Huber}, D., {Sharma}, S., {et~al.} 2015, \apjl, 809, L3

\bibitem[{{Unno} {et~al.}(1989){Unno}, {Osaki}, {Ando}, {Saio}, \&
  {Shibahashi}}]{1989nos..book.....U}
{Unno}, W., {Osaki}, Y., {Ando}, H., {Saio}, H., \& {Shibahashi}, H. 1989,
  {Nonradial oscillations of stars}

\bibitem[{{White} {et~al.}(2011){White}, {Bedding}, {Stello},
  {Christensen-Dalsgaard}, {Huber}, \& {Kjeldsen}}]{White2011}
{White}, T.~R., {Bedding}, T.~R., {Stello}, D., {et~al.} 2011, \apj, 743, 161

\end{thebibliography}
\end{document}